\documentclass{article}
\usepackage{hiph-art}
\usepackage{graphicx}
\usepackage{amssymb}
\volnumber{??} \issuenumber{?} \edyear{200?}                             
\frompage{???} \topage{???}                                              
\recrevdate{29 September 2005}                                           
\def\rightmark{Incomplete thermalization at RHIC}
\def\leftmark{N. Borghini}

\newcommand{\mean}[1]{\langle #1 \rangle}
\newcommand{\Ma}{M{\scriptstyle\!}a}
\newcommand{\Kn}{K{\scriptstyle\!}n}

\title{Hints of incomplete thermalization in RHIC data}
\authors{ 
{Nicolas Borghini\index{Borghini, N.}}\\[2.812mm]
{\normalsize
Physics Department, Theory Division, CERN, CH-1211 Geneva$\,$23, Switzerland}}
 
\abstract{The large elliptic flow observed in Au--Au collisions at RHIC is often
put forward as a compelling evidence for the formation of a strongly-interacting
quark-gluon plasma. 
The main argument is that the measured elliptic flow is as large as the value 
given by fluid-dynamics models that assume complete thermalization. 
It is argued that this claim may not be justified, since a detailed examination 
of experimental data rather suggests that the system created is not fully 
equilibrated at the time when anisotropic flow develops.}


 
\makeindex
\begin{document}
 
\maketitle
\setcounter{page}{1}

\section{Introduction}
\label{s:introduction}

One of the salient results of the heavy-ion programme at RHIC is the measurement
with unparalleled detail and accuracy of the anisotropy in the 
transverse-momentum distributions of particles emitted in the collisions, the 
so-called ``anisotropic flow''.
The four Collaborations have provided plenty of data on the first harmonics 
$v_1$ (``directed flow''), $v_2$ (``elliptic flow'') and $v_4$ in the Fourier 
expansion of the azimuthal distribution of particles, as a function of the 
particle transverse momentum and rapidity, for various particle 
species~\cite{Adler:2003kt,Back:2004zg,Adams:2004bi,Ito:2005}.
These data triggered immense interest, as it was claimed that for the first time
they could be reproduced, together with the transverse-momentum spectra of 
identified particles, by hydrodynamical models~\cite{Kolb:2003dz}. 
Such an agreement supposedly necessitates that the matter created in the 
collisions be thermalized after about 0.6~fm/$c$, and that its equation of state
be soft. 

The claim led to a huge amount of theoretical studies which investigate whether
such a short thermalization time can be accounted for in microscopic models of 
the collision~\cite{TheseProceedings}.
Meanwhile, more phenomenological studies are still needed, to question the 
uniqueness of the interpretation of the experimental findings.
It actually turns out that there is ample room for an alternative reading of 
the anisotropic-flow data, provided one assumes that equilibration in the 
collisions is {\em incomplete\/}~\cite{Bhalerao:2005mm}. 
In that view, I shall first recall in Sec.~\ref{s:hydro-flow} various definite
predictions of ideal fluid dynamics regarding anisotropic flow.
The contrasting predictions of an out-of-equilibrium scenario will then be 
presented in Sec.~\ref{s:nonhydro}; 
in particular, it will be shown that the latter assumption elucidates in a 
natural way several features of the data that were left aside by the ideal-fluid
explanation.

\section{Anisotropic flow in ideal fluid dynamics}
\label{s:hydro-flow}

In this Section, I shall list different firm predictions for anisotropic flow 
derived within ideal fluid dynamics. 
Before that, let me first briefly recall the physical prerequisites for using a 
hydrodynamical description of the evolving matter in heavy-ion collisions. 

\subsection{Ideal fluid dynamics: the basic physics ingredients}

In fluid dynamics, it is customary to sort fluid flows into different categories
according to their physical properties, using dimensionless numbers.
Thus, viscous (resp. inviscid, also referred to as ``ideal'') flows are 
characterized by small (resp. large) Reynolds numbers 
$Re\equiv\varepsilon L\,v_{\rm fluid}/\eta$, where $\varepsilon$, 
$v_{\rm fluid}$ and $\eta$ are the energy density, velocity and shear viscosity 
of the fluid, and $L$ some characteristic length in the system.
Similarly, the Mach number $\Ma\equiv v_{\rm fluid}/c_s$, where $c_s$ is the 
speed of sound in the fluid, quantifies the difference between incompressible 
($\Ma\ll 1$) and compressible ($\Ma\sim 1$) flows.
Finally, the Knudsen number $\Kn\equiv\lambda/L$ --- where $\lambda$ is a mean 
free path --- marks the disparity between systems with low numbers of collisions
per particle (large $\Kn$), which behave like free-streaming gases, and 
liquid-like systems in which each particle experiences many collisions 
($\Kn\ll 1$).

The three above-mentioned numbers are actually related to each other: since 
$\eta=\varepsilon c_s\lambda$, one finds at once $\Ma=\Kn\times Re$.
Now, the matter created in heavy-ion collisions expands into the vacuum, hence 
the corresponding flow is compressible and $\Ma$ is of order unity.
This implies that the expansion of the created fluid satisfies the relation 
$Re\simeq 1/\Kn$: 
if one can show that $\Kn$ is small, it means that the fluid viscosity is small,
in accordance with the ``ideal liquid'' paradigm~\cite{Shuryak:2004cy}.

In the following, I shall present results on anisotropic flow in both regimes 
where $\Kn\sim 1$ (Sec.~\ref{ss:nonhydro-flow}) and $\Kn\ll 1$ 
(Sec.~\ref{ss:hydro-flow}).
In the latter case, which will be referred to as the ``ideal fluid'', the large 
number of collisions per particle leads to some local thermal equilibrium. 
One distinguishes several types of equilibria: either ``kinetic'', i.e., 
equilibrium with respect to {\em elastic\/} collisions, or ``chemical'', with 
respect to {\em inelastic\/} interactions.
It is important to keep in mind that these two equilibria do not necessarily 
hold simultaneously. 
They are also probed by different observables, namely the relative abundances of
particle species for chemical equilibrium, while the constrains on momentum 
distributions imposed by kinetic equilibrium are rather investigated with the 
help of anisotropic flow and HBT correlations~\cite{Borghini:2003nb}.
Unless stated explicitly otherwise, any reference to equilibrium or 
equilibration in the remainder of this paper will actually only concern kinetic 
equilibrium.

\subsection{Predictions for anisotropic flow}
\label{ss:hydro-flow}

Following the original predictions in Ref.~\cite{Ollitrault:1992bk}, the use of 
hydrodynamics provides a simple intuitive picture for the physics of anisotropic
flow.
The initial spatial anisotropy in the transverse plane of the overlap zone 
between two nuclei in a non-central collision results in a stronger pressure 
gradient in the direction of impact parameter (``in-plane'') than perpendicular 
to that direction (``out-of-plane''). 
As a consequence, in-plane particles acquire more momentum than out-of-plane 
particles, leading to an anisotropy of the transverse-momentum distributions.

\begin{figure}
\centerline{\includegraphics*[width=0.75\linewidth]{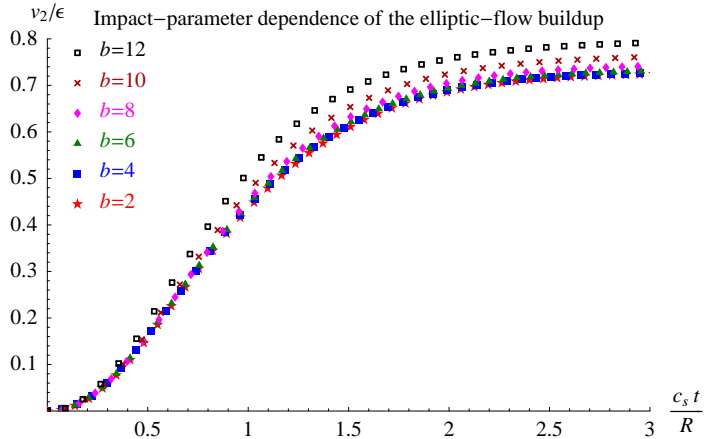}}
\caption{Time evolution of $v_2/\epsilon$ for different values of the impact 
  parameter, for a gas of massless particles ($c_s=1/\sqrt{3}$).
  \label{fig:v2_vs_b}}
\vspace{-3mm}
\end{figure}

More quantitative statements can be made, based either on analytical 
calculations or on Monte-Carlo computations.
Thus, simulations of the development of the average elliptic flow $v_2$ show the
existence of various scalings~\cite{Bhalerao:2005mm}.
As a first example, the time development of $v_2$ is independent of the 
centrality of the collision: if one scales $v_2$ by the initial spatial 
eccentricity $\epsilon\equiv\mean{y^2-x^2}/\mean{y^2+x^2}$ and studies how it 
evolves with time measured in units of $\bar R/c_s$, where 
$\bar R=(1/\mean{x^2}+1/\mean{y^2})^{-1/2}$ quantifies the size of the overlap 
region, one finds a universal curve for most values of the impact parameter, 
except for the most peripheral collisions (see Fig.~\ref{fig:v2_vs_b}).
This universal behaviour shows in particular that the typical buildup time for 
elliptic flow is $\bar R/c_s$, i.e., about 2--4~fm/$c$ for Au--Au or Cu--Cu 
collisions, in agreement with the findings of transport-model
computations~\cite{Bravina:2004td,Chen:2005mr}.

While Fig.~\ref{fig:v2_vs_b} shows that the final $v_2$ value depends on the 
shape of the overlap region, $v_2\propto\epsilon$, the scale invariance of ideal
fluid dynamics implies that it is independent of the system size, i.e., of 
$\bar R$. 
Note, however, that this system-size invariance does not allow a straightforward
extrapolation from one system to the other (say from Au--Au to Cu--Cu 
collisions), because the initial conditions do not scale accordingly (different 
nuclei have different density profiles). 

For given system size and shape, the final elliptic-flow value varies with the 
speed of sound $c_s$. 
More quantitatively, if one assumes a constant $c_s$ throughout the system 
evolution, then the final $v_2$ increases with $c_s$ as soon as $c_s>0.1$ --- it
even becomes  proportional to $c_s$ for $c_s\gtrsim 0.3$~\cite{Bhalerao:2005mm}.

To close this Section, let me mention a few results that were obtained 
analytically, exploiting the fact that the ideal-fluid assumption is equivalent 
to considering the limit of small freeze-out temperature. 
In Ref.~\cite{Borghini:2005kd}, it was emphasized that emitted particles fall 
in a natural way into two categories, namely ``slow'' particles, defined as 
those whose velocity equals that of the fluid at some point on the freeze-out
hypersurface, and ``fast'' particles, which are faster than the fluid at 
freeze-out. 
For both categories of particles, definite predictions regarding anisotropic 
flow can be made. 
Thus, the dependence of elliptic flow on the particle velocity $v_2(p_T/m)$, 
where $p_T$ and $m$ are the particle transverse momentum and mass, should be 
identical for all types of slow particles (except pions, whose mass is not much 
larger than the freeze-out temperature).
The same property holds for all other flow harmonics $v_n(p_T/m)$.
This in particular implies a mass-ordering of $v_2(p_T)$, the heavier particles
having smaller flow at a fixed transverse momentum. 
It turns out that the mass-ordering of $v_2(p_T)$ also holds for fast particles.
For the latter, it was shown that the various even flow harmonics are related, 
the most important relation being $v_4(p_T)=v_2(p_T)^2/2$, valid for each 
particle species in each rapidity window provided the transverse-velocity 
profile at freeze-out is not too different from an 
ellipse~\cite{Borghini:2005kd}.

\section{Out-of-equilibrium scenario}
\label{s:nonhydro}

Let me now turn to the case in which the mean number of collisions per particle,
$\Kn^{-1}$, is insufficient to lead to any (local) equilibrium. 

\subsection{Anisotropic flow in the out-of-equilibrium regime}
\label{ss:nonhydro-flow}

Determining the precise dependence of anisotropic flow on $\Kn^{-1}$ (as well as
on the collision cross sections) requires a transport model which is beyond the 
scope of the present study, yet general predictions are nonetheless 
feasible~\cite{Bhalerao:2005mm}.

\begin{figure}
\centerline{\includegraphics*[width=0.6\linewidth]{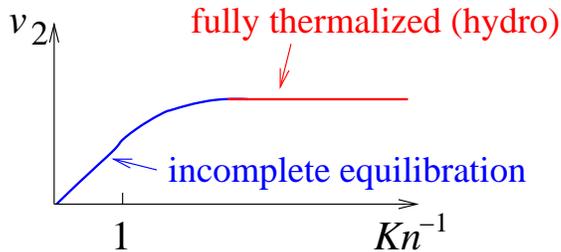}}
\caption{Sketchy representation of the variation of $v_2$ with the mean number 
  of collisions per particle $\Kn^{-1}$.\label{fig:v2_vs_Kn}}
\vspace{-3mm}
\end{figure}

Thus, it is natural to expect that elliptic flow increase with the number of 
collisions, since $v_2$ vanishes in the absence of final-state interaction,
whereas in the large-$\Kn^{-1}$, ideal-fluid limit $v_2$ is finite. 
This growth, however, eventually saturates for some $\Kn^{-1}$ value, for which 
the system equilibrates (see Fig.~\ref{fig:v2_vs_Kn}; ideal-fluid dynamics is 
expected to yield the maximum $v_2$, as hinted at by transport computations, 
which always give smaller values~\cite{Molnar:2004yh}).
The value at which $v_2$ saturates obviously depends on the system shape, i.e., 
on the initial eccentricity $\epsilon$, since $v_2\propto\epsilon$ in 
hydrodynamics; however, the corresponding value of the Knudsen number should be 
quite independent of $\epsilon$, as individual parton--parton or hadron--hadron 
interactions do not know anything about the geometry of the nucleus--nucleus 
collision.
As a consequence, the slope in the region where $v_2(\Kn^{-1})$ grows should be 
roughly proportional to $\epsilon$;
conversely, the ratio $v_2/\epsilon$ should be (almost) independent of the 
centrality of the collision for a fixed value of $\Kn^{-1}$.

The same reasoning applies to $v_4$, which should also increase with the mean 
number of collisions, and then saturate (at a value $\propto\epsilon^2$). 
If $v_4$ does not grow much faster than $v_2$, then the ratio $v_4/v_2^2$ is a 
decreasing function of $\Kn^{-1}$, which reaches a {\em minimum\/} in the 
ideal-fluid limit. 
Since the hydrodynamical prediction for the ratio is $1/2$, then $v_4>v_2^2/2$ 
when equilibrium is not reached~\cite{Bhalerao:2005mm}. 

In the out-of-equilibrium regime, both $v_2$ and $v_4$ increase with the number 
of collisions.
Now, $\Kn^{-1}=L/\lambda$ may vary for two reasons, due to changes either in the
system size $L$, for which a natural choice is $L=\bar R$, or in the mean free 
path $\lambda$. 
The former possibility strongly contrasts with the scale invariance of 
anisotropic flow within the ideal-fluid description~\cite{Bhalerao:2005mm}.
In turn, the mean free path $\lambda=1/n\,\sigma$ depends on both particle 
density $n$ and cross section $\sigma$, and any change in one of these will 
affect the Knudsen number, hence the flow coefficients when the system is not 
equilibrated.

To summarize this part, in an out-of-equilibrium scenario the anisotropic flow 
coefficients $v_2$, $v_4$ depend significantly on the system size $\bar R$ (even
if the system shape $\epsilon$ is fixed), on the particle density and on the 
interaction cross section, and they are related by $v_4/v_2^2>1/2$, the ratio 
increasing as one goes further from equilibrium.

\subsection{Confronting RHIC data and the out-of-equilibrium scenario}
\label{ss:data&non-hydro}

Let me now show that available flow data support the idea that the system 
created in Au--Au collisions at RHIC is not equilibrated, at least at the time 
when anisotropic flow develops.

The first element that supports the out-of-equilibrium scenario is its ability 
to explain the rapidity dependence of elliptic flow~\cite{Hirano:2001eu,%
  Heinz:2004et}.
Thus, the fact that $v_2(y)$ follows closely the rapidity distribution $dN/dy$ 
from midrapidity up to the forward regions, where both exhibit the ``limiting 
fragmentation'' property across different beam energies~\cite{Back:2004zg}, is 
naturally explained within a model where $v_2$ varies with the number of 
collisions per particle.
Now, the identity $(c\bar R/c_s)\,n(\bar R/c_s) = (1/S)\,dN/dy$ yields the 
particle density at the time when anisotropic flow builds up, which gives 
\begin{equation}
\label{Kn-1}
\Kn^{-1} = \bar R \sigma\,n\!\left(\frac{\bar R}{c_s}\right) = 
\frac{c_s}{c}\frac{\sigma}{S}\frac{dN}{dy},
\end{equation}
where $S$ measures the transverse area of the collision zone. 
Equation~(\ref{Kn-1}) shows that the incomplete-equilibration prediction 
$v_2\propto \Kn^{-1}$ translates into $v_2(y)\propto dN/dy$.
On the other hand, the few 3-dimensional hydrodynamical models cannot reproduce 
the data, either overestimating $v_2(y)$ at intermediate rapidities 
$y\sim 3$~\cite{Hirano:2001eu} or, when those values are well predicted, 
underestimating the midrapidity elliptic flow~\cite{Grassi:2005}.

A similar instance of experimental result which finds a natural explanation in 
a non-equilibrium model is the growing discrepancy between $v_2(p_T)$ and the 
ideal-fluid prediction as transverse momentum increases~\cite{Teaney:2003pb}.
As a matter of fact, with increasing $p_T$ the interaction cross section is 
expected to decrease, diminishing the mean number of collisions per particle 
$\Kn^{-1}\propto\sigma$, which is thus increasingly further from the number 
needed to ensure equilibration.
In turn, this leads to an increase of the difference between the hydrodynamical 
$v_2$ and the out-of-equilibrium value. 

One could claim that the previous two arguments concern only a negligible 
fraction of the particles, while the bulk, at midrapidity and moderately low 
transverse momenta, would still be in equilibrium. 
If this were true, then the elliptic flow $v_2$ at midrapidity, averaged over 
transverse momentum, should be proportional to the eccentricity. 
However, it turns out that the ratio $v_2/\epsilon$ is not constant across 
centralities in Au--Au collisions~\cite{Adams:2004bi}: the data do not exhibit 
the scale invariance of ideal fluid dynamics.
On the contrary, the values of $v_2/\epsilon$ rather seem to be scaling linearly
with the control parameter $(1/S)\,dN/dy$~\cite{Alt:2003ab}, i.e., according to 
Eq.~(\ref{Kn-1}) and assuming that $c_s$ and the cross section remain roughly 
constant for the various centralities, with the number of collisions.
This proportionality, without a single hint at any saturation, even extends down
to the values measured in Pb--Pb collisions at the CERN SPS, lending further 
credence to the out-of-equilibrium scenario. 

Yet another indication that the system created in Au--Au collisions at RHIC is 
not equilibrated is provided by the ratio $v_4/v_2^2$. 
STAR and PHENIX reported values which are significantly larger than the 
ideal-fluid value of $1/2$~\cite{Adams:2004bi,Masui:2005}. 

As a final evidence that approaches based on equilibrium assumptions may not be 
appropriate at RHIC, let me point out another failure of the ideal-fluid 
description, which can be resolved provided one drops some equilibrium 
constraint.
Thus, although hydrodynamics describes properly the transverse-momentum 
dependence of $v_2$ of identified particles in {\em minimum bias\/} collisions, 
on the other hand it misses the centrality dependence of $v_2(p_T)$:
the ideal-fluid calculation that reproduce the minimum-bias average overestimate
elliptic flow in peripheral events and underestimate its value for pions in 
central collisions~\cite{Adams:2004bi}.
Even though the former discrepancy is not really surprising --- one does not 
expect the hydrodynamical description to hold for peripheral collisions, where 
the system size is too small to allow any equilibration ---, however the fact 
that data largely {\em overshoot\/} the so-called ideal-fluid limit in central 
events is a much more serious issue. 
A plausible explanation could be that the analysis of the data for these 
centralities is not fully reliable, which is possible since this is where the 
anisotropic-flow signal is smallest, hence most difficult to extract accurately.
Another possibility is that the central data could indeed be described by ideal 
fluid dynamics, albeit with a stiffer equation of state, i.e., a larger speed 
of sound, than what is currently used. 
This can be done, provided one realizes that the supposedly strong constraint 
on $c_s$ arising from fits to rapidity distributions exists if, and only if, 
one assumes that the system has reached not only kinetic equilibrium, but also 
{\em chemical\/} equilibrium. 
Once this assumption (which is only supported by the success of fits from 
``thermal'' models to particle-abundance ratios, as seen also in $e^+ e^-$ 
collisions) is dropped, the one-to-one relationship between particle number 
density, which gives the particle distribution, and energy per particle, which 
is responsible for flow, disappears, and the constraint on $c_s$ is lifted. 
Allowing now kinetic equilibrium only, one may describe central collisions with 
a hydrodynamical model --- however, even if it proved valid for central events, 
the ideal-fluid description would not hold in more peripheral collisions.

It could be hoped that a discriminating test between the early-thermalization 
and out-of-equilibrium scenarios would be provided by measurements of 
anisotropic flow in Cu--Cu collisions~\cite{Bhalerao:2005mm}: 
the change in system size would probe the scale invariance of ideal fluid 
dynamics, which is broken in the absence of equilibrium.
Unfortunately, the preliminary $v_2$ results presented at Quark Matter '05 were 
quite inconclusive, as the values of the three different experimental 
Collaborations were not compatible with each other. 
These discrepancies call for further investigations of the possible sources of 
systematic error on the measurements, in particular fluctuations of the signal 
(whether they arise from truly physical effects, or from binning issues) and 
non-flow effects. 
The latter were shown to affect the flow analysis in Au--Au collisions, and 
should be even more important in the smaller Cu--Cu system; 
several new methods of measurement that are free from their bias were 
specifically introduced~\cite{Borghini:2001vi,Bhalerao:2003xf} and could be 
used.
Unfortunately, disentangling non-flow effects from fluctuations of the flow 
itself is not a trivial task~\cite{Miller:2003kd}. 

Measurements of anisotropic flow with great accuracy will also be needed to 
investigate another expected behaviour, namely that the ratio $v_4/v_2^2$ 
increases at high $p_T$ and/or when going away from rapidity, since in both 
these regimes the measured trends of $v_2(p_T)$ and $v_2(y)$ suggest that 
$\Kn^{-1}$ is decreasing. 

Eventually, one can anticipate that anisotropic flow measurements in Pb--Pb 
collisions at $\sqrt{s_{NN}}=5.5$~TeV at the CERN LHC will help confirm the 
out-of-equilibrium interpretation of RHIC data.
Thus, unless the ideal-fluid limit is marginally reached in the most energetic 
central Au--Au events at RHIC, the average $v_2$ at midrapidity, scaled by the 
spatial eccentricity, should be larger at LHC than it is at RHIC. 
Conversely, the ratio $v_4(p_T)/v_2(p_T)^2$ should be smaller, approaching the 
hydrodynamical value of $1/2$.

\section{Conclusion}

In summary, I have shown that dropping the assumption of kinetic equilibrium 
at the time when anisotropic flow develops allows me to describe in a 
satisfactory, consistent manner several features in the flow data that cannot be 
accommodated in ideal-fluid models.
The only instance where a hydrodynamical approach may remain appropriate is in 
central events; however, even though kinetic equilibrium could be reached, the 
constraint of chemical equilibrium has to be abandoned if one is to reproduce 
the $v_2$ data. 

Although they permit a better description of the data, the predictions of the 
out-of-equilibrium scenario presented here are admittedly quite crude, and
deserve further dedicated studies. 
A transport model would allow a more quantitative investigation of how the 
system evolves from an non-equilibrated state to a thermalized one, answering 
various questions as when (for which value of the Knudsen number) does 
anisotropic flow reach the hydrodynamical limit? 
How far are RHIC Au--Au data from this limit?
In that respect, the ratio $v_4/v_2^2$ might be a more sensitive indicator than 
$v_2$ itself, as it seems to be further away from the ideal-fluid prediction.

\section*{Acknowledgments}

I wish to thank the organizers for their invitation to this beautifully planned
Workshop, which allowed many vivid discussions.

\vfill\eject
\end{document}